\documentclass[superscriptaddress,twocolumn,floatfix,showpacs,prb,10pt]{revtex4-1}
\usepackage{graphicx}
\usepackage{amsmath}
\usepackage{amssymb}
\usepackage{bm}
\usepackage{hyperref}
\usepackage{multirow}
\usepackage{color}
\usepackage{amsbsy}
\usepackage[normalem]{ulem}

\newcommand{\pd}{{\phantom\dag}}

\begin{document}

\title{Topological phase transition in a stretchable photonic crystal}

\author{Ehsan Saei Ghareh Naz}
\email{e.saei.ghareh.naz@ifw-dresden.de}
\affiliation{Institute for Integrative Nanosciences, IFW Dresden, Helmholtzstr. 20, 01069 Dresden, Germany}

\author{Ion Cosma Fulga}
\affiliation{Institute for Theoretical Solid State Physics, IFW Dresden, Helmholtzstr. 20, 01069 Dresden, Germany}

\author{Libo Ma}
\affiliation{Institute for Integrative Nanosciences, IFW Dresden, Helmholtzstr. 20, 01069 Dresden, Germany}

\author{Oliver G. Schmidt}
\affiliation{Institute for Integrative Nanosciences, IFW Dresden, Helmholtzstr. 20, 01069 Dresden, Germany}
\affiliation{Material Systems for Nanoelectronics, Chemnitz University of Technology, Reichenhainer Strasse 70, 09107 Chemnitz, Germany}

\author{Jeroen van den Brink}
\affiliation{Institute for Theoretical Solid State Physics, IFW Dresden, Helmholtzstr. 20, 01069 Dresden, Germany}
\affiliation{Institute for Theoretical Physics, TU Dresden, 01069 Dresden, Germany}

\date{\today}
\begin{abstract}

We design a setup to realize tunable topological phases in elastic photonic crystals. Using the Su-Schrieffer-Heeger (SSH) model as a canonical example, we show how a system can be continuously tuned across its topological phase transition by stretching. We examine the setup both analytically and numerically, showing how the phase transition point may be identified from the behavior of bulk modes.
Our design principle is generic as it can be applied to a variety of systems, and enables multiple new theoretical predictions to be experimentally tested by continuously strain-tuning system properties, such as the shape of the bandstructure and the topological invariant. In addition, it allows for cost-effective device fabrication, since a wide range of parameter space can be accessed on a single photonic crystal chip.

\end{abstract}
% \pacs{...}
\maketitle

\section{Introduction}
\label{sec:intro}

While a topological phase of matter is by its nature robust to small changes in system parameters, reaching a parameter regime in which a topological ground state forms can be a very challenging task experimentally.
The properties of a material are highly constrained by its crystal structure and the process used to fabricate it. There are only few examples in which multiple topological invariants can be obtained in the same sample, for instance by continuously changing the magnetic field in the quantum Hall effect.\cite{Klitzing1980} In most other cases, particularly when changing spin-orbit coupling or the size of the band gap is required, selecting favorable parameters usually involves searching for many different material candidates.\cite{Ando2013, Bradlyn2017}

To overcome these intrinsic constraints, one can instead design a system which simulates the behavior of a topological phase with a given set of parameters. There is a growing literature on such \emph{topological simulators}, setups which mimic the behavior of a topological phase while allowing for a wide range of possible parameter values. They
have been demonstrated in a variety of platforms, like ultra-cold atoms,\cite{Mancini2015, Stuhl2015, Meier2016} driven quantum impurities,\cite{Yuan2017} electrical\cite{Imhof2017} or superconducting circuits,\cite{Tan2017} beam splitters\cite{Kitagawa2012} or microwave networks,\cite{Hu2015} but also mechanical,\cite{Susstrunk2015} acoustic,\cite{Xiao2015, He2016} and optical meta-materials.\cite{Fang2012, Rechtsman2013}
In some cases, the first experimental observation of new types of topological phases was made using simulators. These include higher-order topological insulators,\cite{Imhof2017, Peterson2017} certain types of topological semimetals,\cite{Tan2017} Hopf insulators,\cite{Yuan2017} or so-called anomalous Floquet topological insulators.\cite{Maczewsky2017, Mukherjee2017}

Beyond the advantage of reaching previously unaccessible parameter regimes, some quantum simulators also allow for system properties to be continuously tuned. This can be achieved either by externally applied electric or magnetic fields,\cite{Figotin1998, Chen2017} or by considering the intrinsic deformations of the metamaterial.\cite{Kim2001, Lin2011, Park2004, Wang2013, Liu2018, Lustig2017} For instance, it was recently proposed that applying an electric field can change the refractive index of a photonic crystal, thereby switching between two topologically distinct phases.\cite{Shalaev2018} Similarly, by using flexible components, different topological phases may be reached by applying pressure to the system.\cite{Chen2016, Liu2018} The ability to dynamically change system parameters leads not only to reduced fabrication costs, since multiple regimes of operation can be reached in the same sample, but also facilitates the study of certain peculiar features of topological phases. For instance, the study of disorder on topological phase transitions would require fine-tuning the system to its disorder dependent critical point,\cite{Evers2008} which is more easily achieved in a continuously tunable system.

In this work, we consider a novel design for a system in which parameters can take a large range of possible values, but can also be actively and continuously tuned in the same sample. We focus on a photonic crystal realized as an array of coupled optical waveguides, which simulates the dynamics of electronic wavefunctions in a tight-binding model.\cite{Szameit2010} Each optical waveguide represents a site of the tight-binding model, while the distance between adjacent waveguides sets the strength of the hopping processes. Our main insight is that fabricating the system using a combination of both rigid and elastic materials allows to \emph{selectively} tune desired system properties. Using the Su-Schrieffer-Heeger (SSH) model\cite{Su1979} as a canonical example of a topological phase, we show how our design allows to continuously tune across the topological phase transition. We demonstrate how the precise position of the phase transition in parameter space may be identified by examining the behavior of bulk states.

The rest of our work is organized as follows. In Section \ref{sec:ssh} we briefly review the SSH model, its topological invariant and protected zero-energy end states. We show that the topological phase transition in this model can be identified as a point where bulk modes have a maximal velocity. In Section \ref{sec:pc} we introduce a realistic photonic crystal design that is achievable with current experimental techniques. We numerically analyze the expected experimental signatures and show that the system can be tuned across a topological phase transition by stretching. Finally, in Section \ref{sec:conc} we conclude by discussing how our design strategy may be used in other topological phases, and comment on which properties of topological models may be probed in a continuously tunable system.

\section{SSH model}
\label{sec:ssh}

One of the earliest models of a topological phase is the SSH model,\cite{Su1979} originally introduced to describe polyacetylene. The model consists of one species of spinless fermions hopping on a one-dimensional lattice with staggered nearest-neighbor hopping amplitudes. Using $\hbar=1$ and a lattice spacing $a=1$ throughout the following, the Hamiltonian is:
\begin{equation}\label{eq:ssh}
 H=\sum_i t(1-\delta) c^\dag_{i,A} c^\pd_{i,B} + t(1+\delta) c^\dag_{i,B} c^\pd_{i+1,A} + \rm{h.c},
\end{equation}
where $c^\dag$ and $c$ are fermion creation and annihilation operators, $A$ and $B$ label the sublattices of the chain, and $i$ indexes the unit cells. The average hopping amplitude is $t$, while $\delta\in[-1,1]$  sets the strength and sign of the dimerization. In momentum space, the Hamiltonian reads
\begin{equation}\label{eq:sshk}
 {\cal H}(k) = [t(1 - \delta) + t(1 + \delta) \cos(k)] \sigma_x - t(1 + \delta) \sin(k) \sigma_y,
\end{equation}
with $\sigma_i$ Pauli matrices acting on the sublattice degree of freedom.

\begin{figure}[tb]
 \includegraphics[width=\columnwidth]{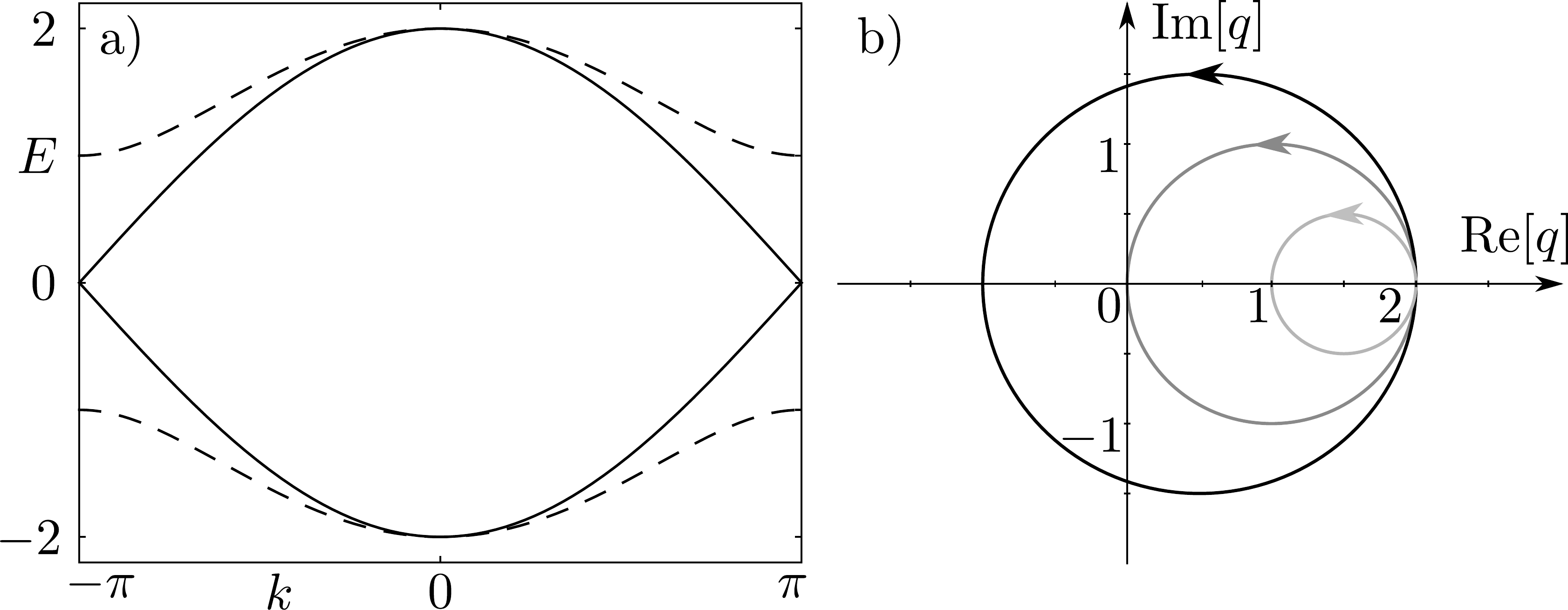}
 \caption{(a): the bandstructure of the SSH chain Eq.~\eqref{eq:sshk} using $t=1$ is plotted for $\delta=0$ (solid lines) and for $|\delta|=0.5$ (dashed lines). The bandstructure is symmetric with respect to $\delta\rightarrow -\delta$. (b): the off-diagonal block of Eq.~\eqref{eq:sshq}, $q(k)$, is plotted in the complex plane for $\delta=0.5$, $0$, and $-0.5$, using again $t=1$. Darker shades indicate larger values of $\delta$ and the arrows indicate the winding direction as $k$ is increased. When the contour encircles the origin, the winding number Eq.~\eqref{eq:windingnr} is nonzero, signaling the presence of zero-energy end states.\label{fig:ssh_toymodel}}
\end{figure}

The bulk spectrum is gapped for $\delta\neq0$ (see Fig.~\ref{fig:ssh_toymodel}a), showing different topological phases depending on its sign. In a long but finite-sized system, there are zero-energy end modes whenever $\delta>0$, such that the chain is terminated with a weak hopping, whereas no end modes appear for $\delta<0$, marking a topologically trivial phase. The presence and robustness of the boundary modes can be inferred from the topology of the bulk states through a topological invariant. The latter is enabled by the presence of sublattice symmetry: there are only two Pauli matrices in Eq.~\eqref{eq:sshk}, such that the Hamiltonian anti-commutes with $\sigma_z$. In this basis, ${\cal H}(k)$ takes a block off-diagonal from
\begin{equation}\label{eq:sshq}
 {\cal H}(k)=\begin{pmatrix}
              0 & q(k) \\
              q^\dag(k) & 0 \\
             \end{pmatrix},
\end{equation}
with $q(k)=t(1-\delta) + t(1+\delta)e^{i k}$. The bulk topological invariant is then given by the winding number of the off-diagonal block:\cite{Asboth2016}
\begin{equation}\label{eq:windingnr}
 \nu = \frac{1}{2\pi i} \int_{-\pi}^{\pi} dk \frac{d}{dk} \log q(k),
\end{equation}
where the integral runs over the entire one-dimensional Brillouin zone.  Whenever $q(k)$ encircles the origin of the complex plane as $k$ is advanced from $-\pi$ to $\pi$ (see Fig.~\ref{fig:ssh_toymodel}b), the integer $\nu$ is nonzero and topologically protected zero modes appear at the ends of a finite chain. As long as the system shows sublattice symmetry, the Hamiltonian can always be made off-diagonal as in Eq.~\eqref{eq:sshq} and $q(k)$ is well-defined. The winding number is left invariant by variations of the system parameters which do not close the bulk gap, since $\nu$ and can only change if $q=0$ for some value of momentum, which implies a gapless system, ${\cal H}=0$.

In a generic SSH chain, which may include more than two bands as well as longer range hopping processes, a topological phase transition should be identified by using the topological invariant Eq.~\eqref{eq:windingnr}. The latter may be determined experimentally from the long-time dynamics of bulk modes, by computing the so-called \emph{mean chiral displacement},\cite{Cardano2017, Maffei2018} as we discuss in the next section.
However, preluding on the numerical results on the topological phase transition in the stretchable photonic system, we can identify in the nearest neighbor SSH model Eq.~\eqref{eq:sshk} a different indicator of the phase transition.
As we will show in the following, the change of the winding number occurs at a point at which bulk modes have maximal velocity. Without loss of generality, we set $t=1$ in the following, and determine the eigenvalues of Eq.~\eqref{eq:sshk},
\begin{equation}\label{eq:evals}
 E_\pm(\delta,k) = \pm \sqrt{2}\sqrt{(1+\delta^2) + (1-\delta^2)\cos(k)},
\end{equation}
as well as the associated velocities of the states,
\begin{equation}\label{eq:vel}
 v_\pm(\delta, k)=\frac{\partial E_\pm(\delta,k)}{\partial k}= \frac{(\delta^2-1)\sin(k)}{E_\pm (\delta, k)}.
\end{equation}

Due to the simple form of the above equations it is possible to analytically determine the maximal velocity as a function of dimerization. We find that the largest velocity has a magnitude $v_{\rm max} = 1-|\delta |$, and occurs at momenta 
\begin{equation}\label{eq:kmax}
k_{\rm max} = \pm {\rm arccos} \left( \frac{|\delta|-1}{|\delta|+1} \right).
\end{equation}

As expected from the bandstructure of Fig.~\ref{fig:ssh_toymodel}a, the maximal velocity occurs at the topological phase transition, corresponding to the linearly dispersing states at the gap closing point, $E=0$ and $k=\pi$. As $|\delta|$ is increased away from $0$, the bands become progressively flatter such that the velocity decreases, eventually vanishing in the fully dimerized flat-band limit, $\delta=\pm1$. While the maximum velocity of all states is achieved at specific momenta $k_{\rm max}$ in Eq.~\eqref{eq:kmax},
the bulk state velocity Eq.~\eqref{eq:vel} is a monotonically decreasing function of $|\delta|$ for any value of momentum. Therefore, in this simple model the position of the phase transition may be determined by examining how a reference bulk state spreads as a function of time, and identifying the point at which this spread is maximal.

\section{Photonic crystal}
\label{sec:pc}

As mentioned in the introduction, photonic crystals offer a versatile platform for the simulation of a variety of topological phases.\cite{Fang2012, Rechtsman2013, Maczewsky2017, Mukherjee2017, Arkinstall2017, St-Jean2017} This is due to the fact that the equation describing light propagation in an array of coupled waveguides, the paraxial Helmholtz equation, has a mathematical structure identical to that of the Schr{\"o}dinger equation.\cite{Szameit2010} The refractive index of the photonic crystal plays the role of the quantum-mechanical potential energy, while the direction of light propagation along the waveguides plays the role of time. When the waveguide diameter is comparable to the light wavelength such that it traps a single propagating mode, then the waveguide acts as a single site in an effective tight-binding model. The strength of hopping processes between neighboring sites is determined by the overlap between the evanescent tails of modes in neighboring waveguides, which decays exponentially with their separation.

The mapping between the Helmholtz and Schr{\"o}dinger equations offers several advantages. First, it allows to directly visualize the simulated temporal evolution of wavepackets by monitoring the spatial propagation of light. This avoids the inherent difficulty of resolving time on very short scales in condensed matter systems. Second, it allows to freely select the magnitude of hopping processes between any neighboring sites, which only requires adjusting the distance between waveguides. Thanks to these advantages, the SSH model was successfully simulated in waveguide arrays with an alternating spacing,\cite{Weimann2017, Bleckmann2017} which model the dimerized hopping of Eq.~\eqref{eq:ssh}.

However, if the desired potential is made by carving the refractive index in a rigid propagation medium (for instance using a laser), the topological phase of the structure becomes fixed. Various different devices then have to be fabricated for demonstrating and analyzing different parameter regimes. In Ref.~\onlinecite{Weimann2017} for instance, an SSH model with different values of the dimerization $\delta$ was investigated by building 9 different photonic crystals with different waveguide spacings. If fine-tuned system parameters are required, for instance to produce a system precisely at a topological phase transition, then imperfections in the fabrication process may mean that an even larger number of samples must be produced.

In order to achieve photonic crystals allowing to tune only the parameters of interest, we propose to combine rigid and elastic materials. For instance, if a one-dimensional array of rigid waveguides were placed on an elastic substrate, then stretching the substrate would continuously change the distance between neighboring waveguides. On the level of the effective tight-binding model, this would imply an approximately uniform decrease in the value of all hopping amplitudes. While this might be useful in some applications, it would not help in the SSH chain. Uniform stretching would decrease the value of $t$ in Eq.~\eqref{eq:ssh} without changing $\delta$, merely leading to a decreased bandwidth without changing the winding number.

\begin{figure}[tb]
 \includegraphics[width=0.8\columnwidth]{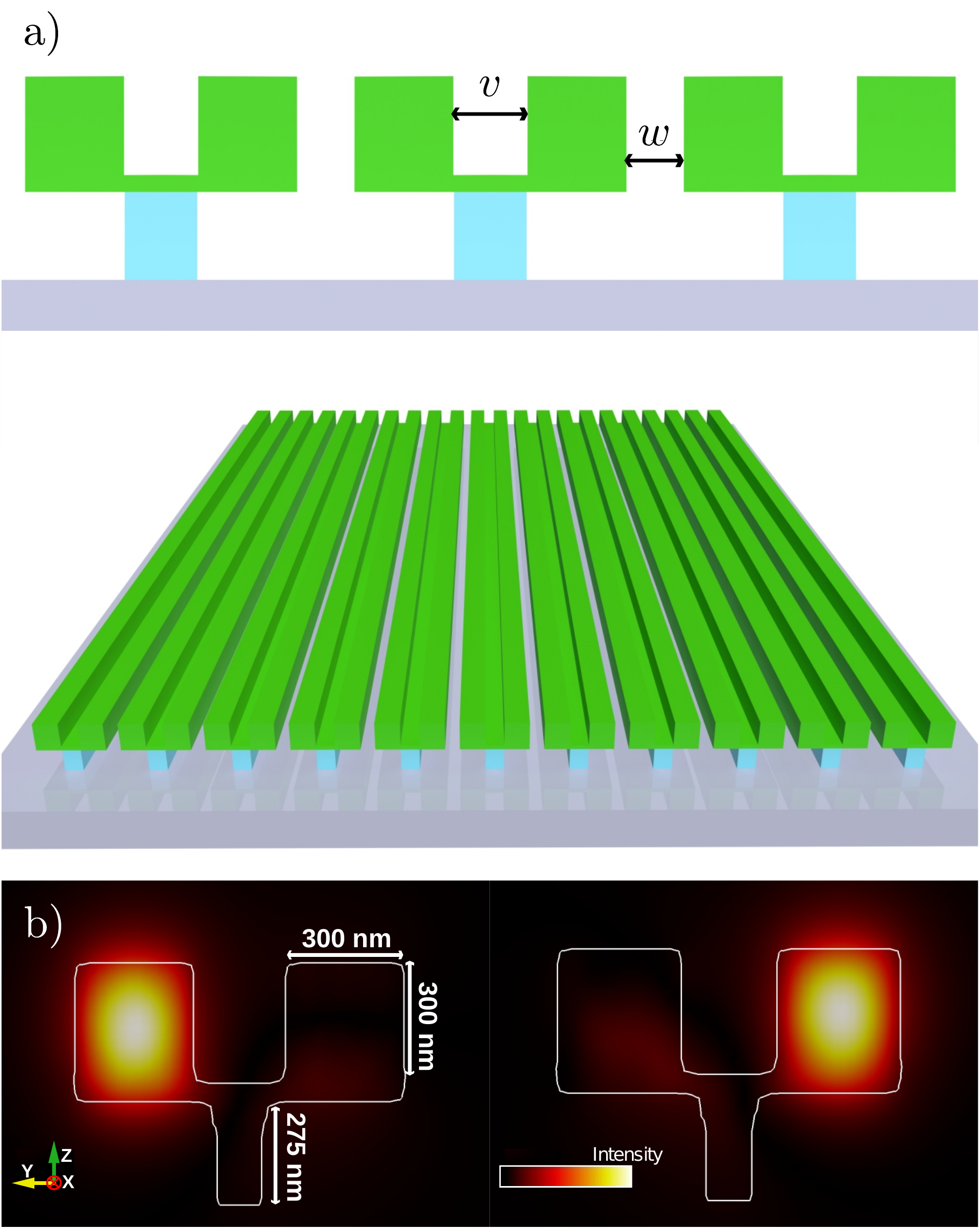}
 \caption{(a): front and perspective view of the stretchable photonic crystal. Pairs of waveguides arranged to form a Y-shaped structure are placed on an elastic substrate, allowing to tune the inter-cell hopping ($w$) while keeping the intra-cell hopping ($v$) fixed. (b): field confinement in the plane perpendicular to the propagation direction at two different positions along the waveguides as calculated numerically by FDTD  simulations (see text). Light is confined within one the the two Y-arms, without leaking to the supporting post. The dimensions of the Y-structure are indicated with arrows.
 \label{fig:setup}}
\end{figure}

To design an SSH chain which allows to change the parameter responsible for the topological phase transition, $\delta$, we propose a system in which the distance between neighboring waveguides can be tuned \emph{selectively}. To this end, we consider that pairs of neighboring waveguides are attached to each other using rigid materials, forming a Y-shaped structure as shown in 
Fig.~\ref{fig:setup}. As such, stretching the substrate will have a negligible effect on the hopping between “connected” sites ($v$), but will lead to a large increase in the separation between disconnected waveguides ($w$).

\begin{figure}[tb]
 \includegraphics[width=0.8\columnwidth]{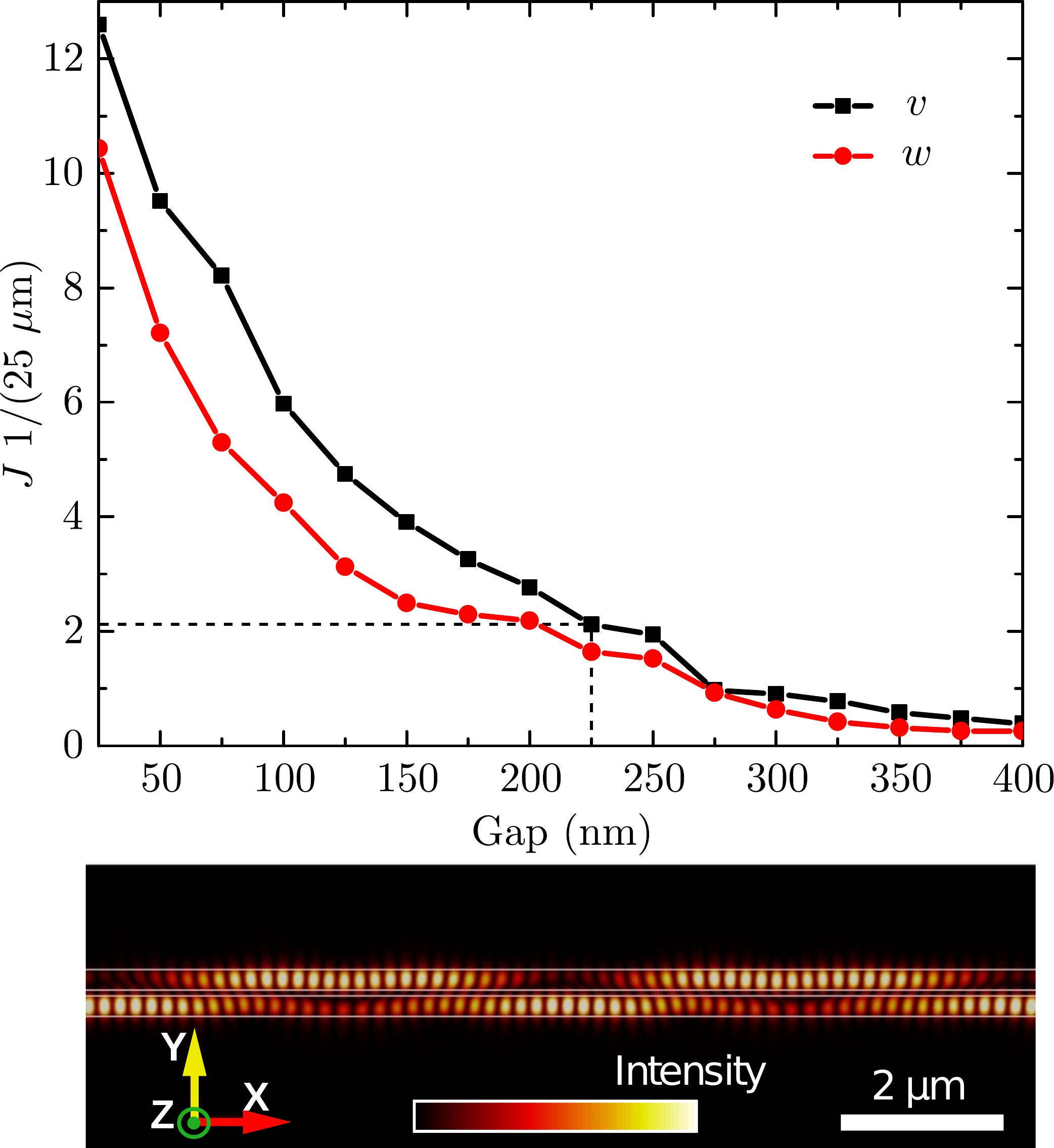}
 \caption{Top: strength of the nearest neighbor hopping as a function of waveguide separation. The strength of intra-cell ($v$) and inter-cell ($w$) hopping are different, reflecting the different media between the waveguides. Dashed lines show the value of $v$ used in subsequent simulations. Bottom: 
oscillation of the light field
in an isolated pair of waveguides, using a separation of 75 nm. The light is confined to each of the waveguides forming the Y-structure and jumps to the neighboring waveguide periodically. The frequency of jumps is used to extract the hopping strength.\label{fig:hopping}}
\end{figure}

\begin{figure}[tb]
 \includegraphics[width=0.8\columnwidth]{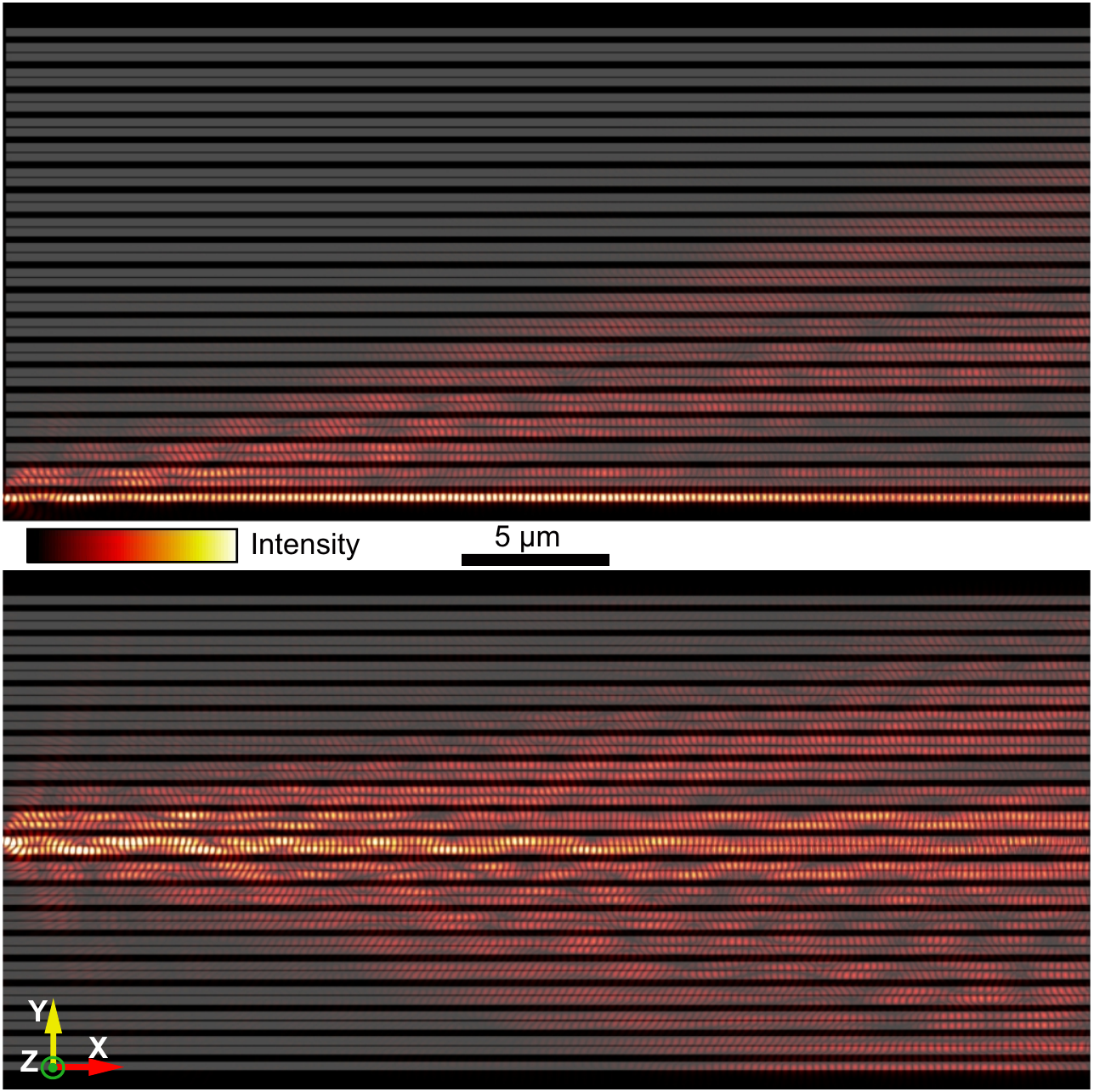}
 \caption{Light intensity profile for the edge (top) and bulk (bottom) modes in the topologically non-trivial phase, corresponding to a separation of 25 nm between waveguides of neighboring Y-structures. The waveguides are arranged horizontally, such that in the simulated SSH model the vertical direction represents space, while the horizontal direction represents time. Most of the light remains confined to the edge, signaling the presence of a topologically protected boundary state, while it disperses in the bulk, owing to the finite velocity of bulk states.\label{fig:edgeandbulk}}
\end{figure}

The Y-shaped pair of waveguides of Fig.~\ref{fig:setup} defines a unit cell of the SSH model, so building a finite-sized chain requires placing multiple equally spaced Ys on an elastic substrate, such as Polydimethylsiloxane. While in principle the only requirement for a tunable dimerization is to have pairs of connected sites, we have chosen a Y-shaped geometry for two reasons. First, the two waveguides of the structure (the arms of the Y) are isolated from each other by an ultra-thin wall ($50$ nm) and also from the substrate through the supporting post, which has a lower refractive index $n$. For instance, one could use Al$_2$O$_3$ for the Y-arms ($n = 1.76$) and SiO$_2$ ($n = 1.45$) for the Y-post. Both the thin connection between the arms and the lower refractive index of the post mean that propagating light will be confined to only one of the waveguides and will not leak to other parts of the structure or to the substrate (see Fig.~\ref{fig:setup}b). As such, the Y forms a well-defined pair of sites in the SSH chain. The second reason for this geometry is that it allows to reach larger values of dimerization. Each Y only touches the substrate on a narrow region, being glued to it thanks to van der Waals forces, such that it remains in place even as the system is stretched. This narrow region
leaves a larger portion of free substrate between neighboring unit cells, so the system can be stretched more easily. Also, since the post is narrow, there is a reduced chance it will detach from the substrate upon stretching.

We test our elastic photonic crystal design by performing Finite Difference Time Domain (FDTD) numerical simulations using the MEEP software package.\cite{Oskooi2010, Naz2017} We use a uniform Yee grid with the size of 25 nm$^3$ and define a photonic crystal of 40 $\mu$m in the longitudinal direction. Each Y-structure has waveguides of size $300\times300$ nm$^2$, connected by a 50 nm thick wall. The width and height of the supporting post are $225$ and $275$ nm, respectively. The refractive indexes are the same as before, $n = 1.76$ for the arms and $n = 1.45$ for the post, corresponding to Al$_2$O$_3$ and SiO$_2$, respectively. In the numerical simulations, we compute the propagation of light (wavelength $\lambda=700$ nm) after it is injected in an initial waveguide. In the longitudinal direction, the waveguides are terminated by a perfectly matched layer boundary condition, such that the light is absorbed at the end of the system and there is no back-reflection.

As a first step, we determine the strength of the hopping processes, both within a unit cell ($v$) as well as between cells ($w$), as a function of waveguide separation. This is done by simulating only a pair of waveguides, and counting how many times light bounces back and forth while propagating for 25 $\mu$m along the waveguides. Results are plotted in Fig.~\ref{fig:hopping}, showing the expected exponential decay. Notice how the two hopping strengths are different for the same separation, reflecting the different media between waveguides. The inter-cell hopping $w$ occurs through air, while the intra-cell $v$ also occurs through the rigid contact between waveguides. Having established the range of parameters available in the simulation, we fix a value $v\simeq 2.11/25\mu$m, corresponding to a separation of 225 nm between the Y-arms. In a system of 19 unit cells, we compute light propagation for different values of $w$, ranging from a separation of 25 nm to 400 nm in steps of 25 nm (the Yee grid size).

Figure \ref{fig:edgeandbulk} shows simulation results for a separation of $25$ nm between waveguides belonging to neighboring Y-structures, such that the system is in a topologically non-trivial phase. When the initial light is sent through only the bottom most waveguide, most of it remains confined to the boundary of the system, reflecting the presence of a topologically non-trivial end mode. In contrast, starting with light in one of the middle waveguides, we observe that it spreads throughout the entire system, since bulk modes have a finite velocity. In both cases, the pattern of light propagation simulates the time evolution of an initial wavepacket localized on either a bulk site or an end site of the SSH chain.

\begin{figure}[tb]
 \includegraphics[width=0.95\columnwidth]{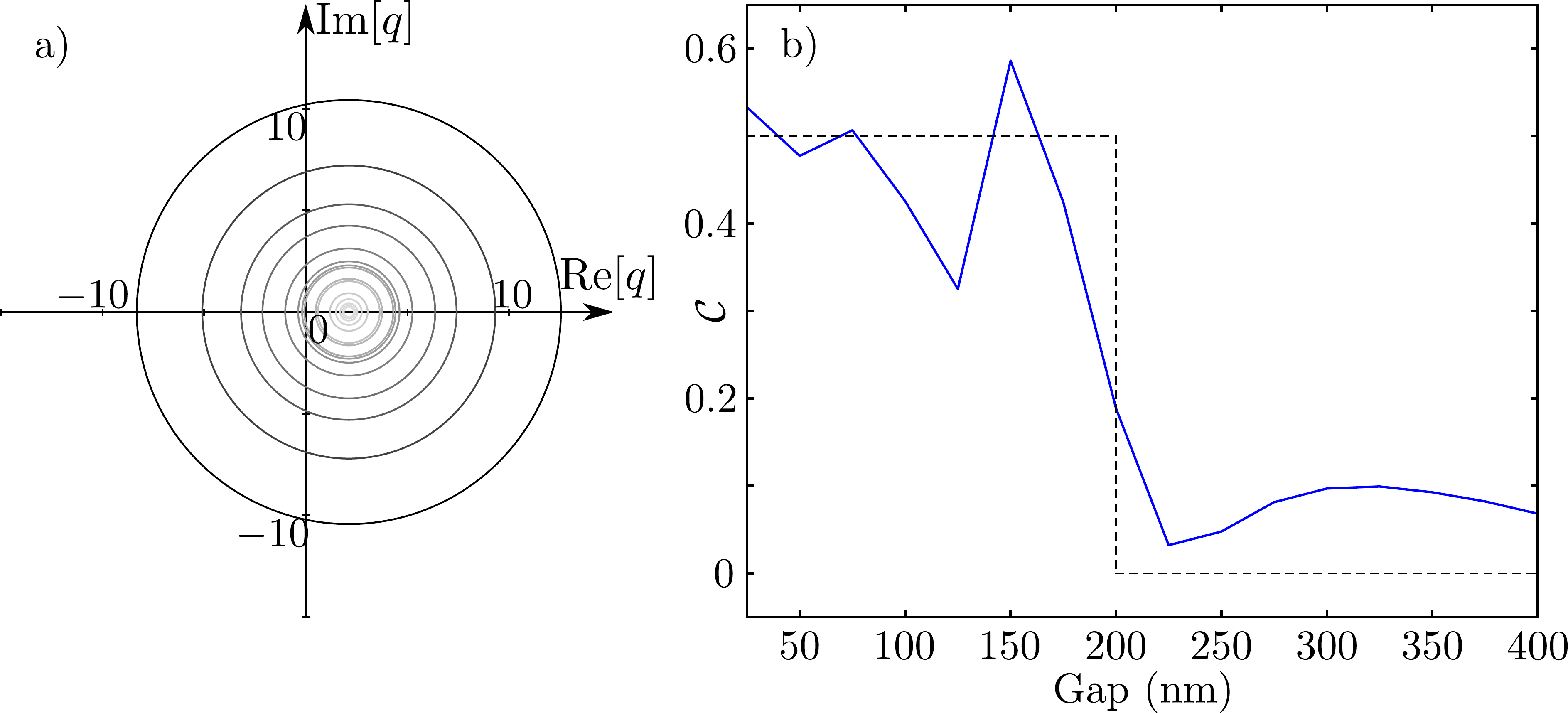}
 \caption{(a): Contours of the off-diagonal block $q(k)$ in Eq.~\eqref{eq:sshq} as momentum is advanced from $-\pi$ to $\pi$, computed using the numerically extracted values of intra-cell and inter-cell hopping, $v$ and $w$. Each contour winds in an anti-clockwise fashion as momentum is increased. Lighter shades and smaller radii correspond to larger separation between the waveguides of adjacent Y-structures (from 25 nm to 400 nm in steps of 25 nm). (b): Mean chiral displacement computed as in Eq.~\eqref{eq:mcd}, plotted as a function of the separation between unit cells. The dashed black line shows the result expected in the infinite time limit. As the photonic crystal is stretched, a topological phase transition to a trivial phase takes place. The latter occurs when the contours of panel (a) no longer encircle the origin and is marked by a sharp drop in the mean chiral displacement of panel (b), from values close to $0.5$ to values close to $0$.\label{fig:stretch_winding}}
\end{figure}

As the photonic crystal is progressively stretched, the simulated SSH chain undergoes a topological phase transition. The latter is signaled by a closing and reopening of the bulk gap and is accompanied by a change in the winding number Eq.~\eqref{eq:windingnr}. For the values of the inter-cell hopping in the simulated experiment, we plot in Fig.~\ref{fig:stretch_winding}a the contour of the off-diagonal Hamiltonian block $q(k)$ [Eq.~\eqref{eq:sshq}]. Notice that unlike in Fig.~\ref{fig:ssh_toymodel}b, the contours are now concentric since the distance between the Y-arms remains constant throughout the stretching process. Nevertheless, a change in the winding number is clearly visible and occurs at the point where the hopping between waveguides of neighboring Ys becomes smaller than the hopping within a unit cell.

\begin{figure}[tb]
 \includegraphics[width=0.7\columnwidth]{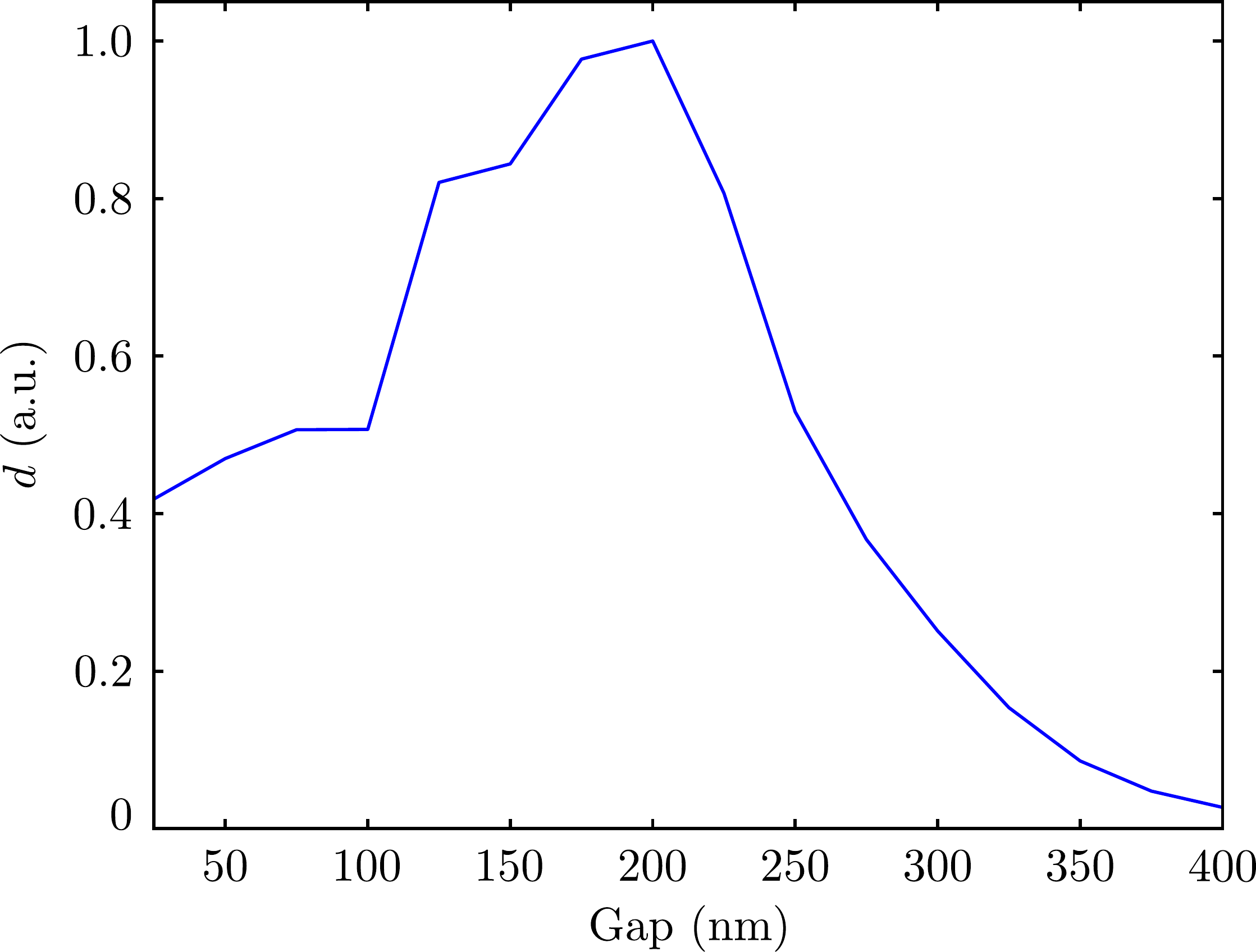}
 \caption{Mean squared displacement $d$ computed as in Eq.~\eqref{eq:msd} and plotted as a function of the distance between unit cells. The value of $d$ has a maximum at the topological phase transition, since at that point the bulk mode velocity is maximal.\label{fig:meanspread}}
\end{figure}

As mentioned in the previous section, the topological invariant can be obtained directly from the light intensity profile by computing the mean chiral displacement.\cite{Cardano2017, Maffei2018} Starting from an initial wavepacket $|\psi\rangle$ localized on a single site in the middle the SSH chain, which we label as belonging to the 0$^{\rm th}$ unit cell, the mean chiral displacement as a function of time reads ${\cal C}(t) = \langle \psi(t) |\Gamma m|\psi(t) \rangle$. Here, $m = {\rm diag} (\ldots, -2, -2, -1, -1, 0, 0, 1, 1, 2, 2, \ldots)$ is an operator labeling the unit cell of each site in the tight-binding model, whereas $\Gamma = \sigma_z \oplus \sigma_z \oplus \ldots \oplus \sigma_z$ is the chiral symmetry operator of the finite chain. In the infinite time limit, it has been shown that ${\cal C}(t)$ converges to value equal to half of the winding number.\cite{Cardano2017, Maffei2018}
To compute the latter, we simulate the propagation of light starting from a single waveguide in the middle of the photonic crystal. The light intensity in each waveguide is integrated over the last 400 nm of the system (corresponding to $1\%$ of the total waveguide length), and the resulting light intensity profile is normalized and used to compute the mean chiral displacement as
\begin{equation}\label{eq:mcd}
 {\cal C} = \sum_j I_{j,A} \cdot j - \sum_j I_{j,B} \cdot j,
\end{equation}
where $I_{j,A/B}$ is the light intensity in the waveguide belonging to the $j^{\rm th}$ unit cell and the $A$ or $B$ sublattice. Results shown in Fig.~\ref{fig:stretch_winding}b confirm that a topological phase transition indeed occurs as a function of stretching. For small distances between neighboring Y-structures, ${\cal C}$ takes values close to $0.5$, indicating a winding number $\nu=1$. As the distance between unit cells is increased, the mean chiral displacement shows a sharp drop to values close to zero, indicative of a topologically trivial phase.

\begin{figure}[tb]
 \includegraphics[width=0.9\columnwidth]{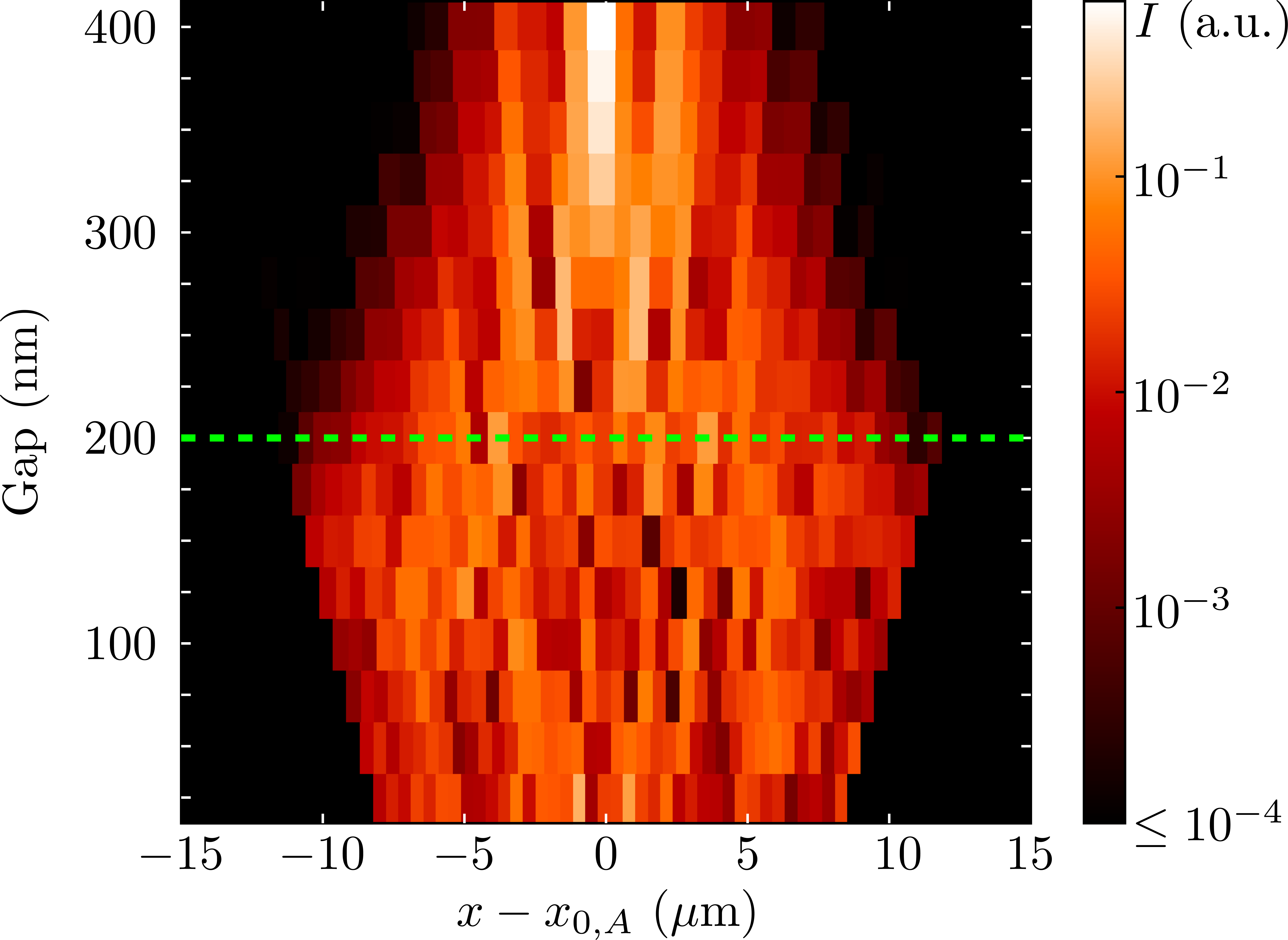}
 \caption{Light intensity profiles after a propagation distance of 40 $\mu$m, starting from a bulk state. The light intensity (color scale) is plotted as a function of waveguide position relative to the initially excited waveguide (horizontal axis) and distance between unit cells (vertical axis). The overall spread of light increases as one approaches the phase transition (dashed green line), and decreases afterwards. The light intensity is plotted on a logarithmic scale to better show the behavior close to the critical point.\label{fig:top_tuning}}
\end{figure}

Beyond directly measuring the topological invariant, the presence and position of a topological phase transition can be more directly observed by examining the spread of light as it propagates through the photonic crystal. As pointed out in the previous section, in this simple model of an SSH chain the average velocity of bulk modes should increase as the critical point is approached, and decrease afterwards. We test this prediction numerically, starting from an initially excited waveguide in the bulk of the photonic crystal. After a distance of $40$ $\mu$m from the light injection point, corresponding to an evolution of wavepackets for a fixed amount of time, we determine the mean squared displacement of light
\begin{equation}\label{eq:msd}
 d = \sum_{j,s} I_{j,s} (x_{j,s} - x_{0,A})^2.
\end{equation}
Here, $I_{j,s}$ is the light intensity in the waveguide belonging to the $j^{\rm{th}}$ unit cell and sublattice $s=A,B$, whereas $x_{j,s}$ is its real space position, which depends on how much the system is stretched. The initially excited waveguide is located in the middle of the system and belongs to the $A$ sublattice, its position being denoted by $x_{0, A}$. Figure \ref{fig:meanspread} shows the normalized value of $d$ determined for different separations between the Y-structures. As expected, the mean squared displacement is maximal at the phase transition point and decreases away from it. Since $d$ is computed after a fixed distance in the photonic crystal lattice, corresponding to a fixed time in the evolution of wavepackets, its maximum value corresponds to a maximum in the bulk state velocity. This non-monotonic behavior in the spread of light through the system can also be directly visualized in the waveguide array, as shown in Fig.~\ref{fig:top_tuning}.

\section{Conclusion}
\label{sec:conc}

We have theoretically proposed and numerically analyzed a new photonic crystal design, which allows to continuously tune system parameters in the same sample. By using an elastic substrate and waveguide pairs arranged in a Y-shaped geometry, we have shown that the lattice constants of the simulated tight-binding model can be addressed selectively. We have tested this design on one of the simplest examples of a topological phase, the SSH model, which can be elastically and reversibly stretched across a topological phase transition. The critical point separating topologically distinct phases can be determined as the separation at which bulk modes have a maximal spread.

Our proposed setup should be accessible in experiments, using currently available fabrication techniques. For instance, the Al$_2$O$_3$ arms of the Y-structure may be carved by electron beam lithography, while the supporting posts made from SiO$_2$ could be produced using wet etching. 
Moreover, we expect that the strategy of combining rigid and elastic materials should be applicable also to other topological models, in both one and two dimensions. For instance, a three-dimensional array of coupled waveguides was recently used to simulate Chern insulators.\cite{Rechtsman2013} In such a system, if one keeps the waveguides rigid but embeds them in an elastic medium, then stretching the photonic crystal should tune the tight-binding model across a two-dimensional topological phase transition.

A single photonic crystal able to reach a wide range of parameter values would mean a reduction in the costs associated to device fabrication. Beyond this practical aspect however, tunable devices would allow experiments to probe a variety of theoretical predictions which would be impractical otherwise. For instance, in photonic crystals realizing topological phases protected by lattice symmetries, so-called topological crystalline insulators,\cite{Ando2015} stretching would allow to probe the robustness of boundary states upon breaking the protecting symmetries. Another example is 
the problem of localization in one-dimensional systems, which has a large, well-established body of theoretical research.\cite{Anderson1958, Theodorou1976, Eggarter1978, Brouwer2000, Motrunich2001, Brouwer2003, Gruzberg2005} While initially it was expected that even an infinitesimal amount of disorder would lead to a localization of all bulk states,\cite{Anderson1958} it was later shown that some states remain delocalized when the one-dimensional system is at a topological phase transition.\cite{Motrunich2001, Gruzberg2005} The critical behavior of a variety of one dimensional systems was analyzed and theoretical predictions for transport properties and the density of states are available, yet the problem has never been tackled experimentally.

\acknowledgments

We thank Ulrike Nitzsche for technical assistance.

\bibliography{ssh}

\end{document}